\documentclass[11pt]{article}
\usepackage[]{amsmath,amssymb,amsfonts,latexsym,amsthm,enumerate,fullpage}
\newtheorem{thm}{Theorem}
\newtheorem*{thm*}{Theorem}
\newtheorem{lemma}[thm]{Lemma}
\newtheorem*{lemma*}{Lemma}

\newtheorem*{prop*}{Proposition}
\newtheorem{cor}[thm]{Corollary}

\newtheorem{conj}[thm]{Conjecture}
\newtheorem{claim}[thm]{Claim}
\theoremstyle{remark}

\newtheorem*{rmk*}{Remark}
\newtheorem*{rmks*}{Remarks}
\newtheorem*{not*}{Notation}
\newtheorem*{claim*}{Claim}
\newtheorem*{fact*}{Fact}

\theoremstyle{definition}
\newtheorem{dfn}{Definition}

\def\E{\mathbb{E}}
\def\F{\mathbb{F}}
\def\P{\mathbb{P}}

\newcommand\ignore[1]{}

\begin{document}

\title{The List-Decoding Size of Reed-Muller Codes}
\author{
Tali Kaufman
\thanks{Research supported in part by
NSF Awards CCF-0514167 and NSF-0729011.}\\
MIT\\
kaufmant@mit.edu\\
\and Shachar Lovett
\thanks{Research supported partly by
the Israel Science Foundation (grant 1300/05). Research was conducted
partly when the author was an intern at Microsoft Research.}\\
Weizmann Institute of Science\\
shachar.lovett@weizmann.ac.il } \maketitle


\begin{abstract}
In this work we study the list-decoding size of Reed-Muller codes.
Given a received word and a distance parameter, we are interested in
bounding the size of the list of Reed-Muller codewords that are
within that distance from the received word. Previous bounds of
Gopalan, Klivans and Zuckerman~\cite{GKZ08} on the list size of
Reed-Muller codes apply only up to the minimum distance of the code.
In this work we provide asymptotic bounds for the
list-decoding size of Reed-Muller codes that apply for {\em all}
distances. Additionally, we study the weight distribution of
Reed-Muller codes.
Prior results of Kasami and Tokura~\cite{KT70} on
the structure of Reed-Muller codewords up to twice the
minimum distance, imply bounds on the weight distribution of the
code that apply only until twice the minimum distance.
We provide accumulative bounds for the weight distribution of Reed-Muller
codes that apply to {\em all} distances.
\end{abstract}

\section{Introduction}\label{sec:intro}

The problem of list-decoding an error correcting code is the
following: given a received word and a distance parameter find all
codewords of the code that are within the given distance from the
received word. List-decoding is a generalization of the more common
notion of unique decoding in which the given distance parameter
ensures that there can be at most one codeword of the code that is
within the given distance from the received word. The notion of
list-decoding has numerous practical and theoretical implications.
The breakthrough results in this field are due to Goldreich and
Levin~\cite{GL} and Sudan~\cite{Sud} who gave efficient list
decoding algorithms for the Hadamard code and the Reed-Solomon code.
See surveys by Guruswami~\cite{Gur} and Sudan~\cite{Sud2} for
further details. In complexity, list-decodable codes are used to perform
hardness amplification of functions~\cite{STV}. In cryptography, list-decodable
codes are used to construct hard-core predicates from one
way functions~\cite{GL}. In learning theory, list decoding of
Hadamard codes implies learning parities with noise~\cite{KM}.


In this paper we study the question of list-decoding Reed-Muller
codes. Specifically, we are interested in bounding the list sizes
obtained for different distance parameters for the list-decoding
problem.

Reed-Muller codes are very fundamental and well studied codes. $RM(n,d)$ is a
linear code, whose codewords $f \in
RM(n,d):\F_2^n \to \F_2$ are evaluations of polynomials in $n$
variables of total degree at most $d$ over $\F_2$.
In this work we study the code $RM(n,d)$ when $d \ll n$,
and are interested in particular in the case of constant $d$.

The following facts regarding $RM(n,d)$ are straight-forward:
It has block length of $2^n$, dimension $\sum_{i \leq d}{n \choose i}$
and minimum relative distance $\frac{2^{n-d}}{2^n} = 2^{-d}$. We define:

\begin{dfn} [Relative weight of a function]
The relative weight of a function/codeword $f:\F_2^n \to \F_2$ is the fraction of non-zero elements,
$$
wt(f) = \frac{1}{2^n} |\{x \in \F_2^n: f(x) = 1\}|
$$
\end{dfn}

A closely related definition is the distance between two functions

\begin{dfn} [Relative distance between two functions]
The relative distance between two functions $f,g:\F_2^n \to \F_2$ is defined as
$$
dist(f,g) = \P_{x \in \F_2^n}[f(x) \ne g(x)]
$$
\end{dfn}

The main focus of this work is in understanding the asymptotic
growth of the list size in list-decoding of Reed-Muller codes, as a
function of the distance parameter. Specifically we are interested
in obtaining bounds on the following.

\begin{dfn}[List-decoding size]
For a function $f:\F_2^n \to \F_2$ let the ball at relative distance $\alpha$ around $f$ be
$$
B(f,\alpha) = \{p \in RM(n,d): dist(p,f) \le \alpha \}
$$

The list-decoding size of $RM(n,d)$ at distance $\alpha$, denoted by
$L(\alpha)$, is the maximal size of $B(f,\alpha)$ over all possible
functions $f$, i.e.
$$
L(\alpha) = \max_{f:\F_2^n \to \F_2} |B(f,\alpha)|
$$
\end{dfn}

In a recent work Gopalan, Klivans and Zuckerman \cite{GKZ08} prove
that for distances up to the minimal distance of the code, the
list-decoding size of Reed-Muller codes remains constant.
\begin{thm}[Theorem 11 in \cite{GKZ08}]\label{thm:gkz2}
$$
L(2^{-d}- \epsilon) \le O \left( (1/\epsilon)^{8d} \right)
$$
\end{thm}

Their result of bounding the list-decoding size of Reed-Muller codes
is inherently limited to
work up to the minimum distance of the code, since it uses a
structural theorem of Kasami and Takura on Reed-Muller codes~\cite{KT70},
which implies a bound on the
weight distribution of Reed-Muller codes that works up to twice the
minimum distance of the code.

Additionally, the work of~\cite{GKZ08} has developed a list-decoding algorithm for $RM(n,d)$ whose running time is polynomial in
the worst list-decoding size and in the block length of the code.
\begin{thm}[Theorem 4 in \cite{GKZ08}]\label{thm:gkz:global-alg}
Given a distance parameter $\alpha$ and a received word $R:\F_2^n \to \F_2$,
there is an algorithm that runs in time
$poly(2^n, L(\alpha))$ and produces a list of all $p \in RM(n,d)$
such that $dist(p,R) \leq \alpha$.
\end{thm}
Since Gopalan et al. could obtain non-trivial bounds on the
list-decoding size for distance parameter $\alpha$ that is bounded
by the minimum distance of the Reed-Muller code, their algorithm
yields meaningful running time only for $\alpha$ that is less than
twice the minimum distance of the code.

\subsection{Weight distribution of Reed-Muller codes}
A close notion to the list-decoding size of Reed-Muller code is the
weight distribution of the code.

\begin{dfn} [Accumulative weight distribution]
The accumulative weight distribution of $RM(n,d)$ at a relative
weight $\alpha$ is the number of codewords up to this weight, i.e.
$$
A(\alpha) = |\{p \in RM(n,d): wt(p) \le \alpha \}|
$$
where $0 \le \alpha \le 1$.
\end{dfn}

It is well-known that for any $p \in RM(n,d)$ which is not
identically zero, $wt(p) \ge 2^{-d}$. Thus, $A(2^{-d} - \epsilon)=1$
for any $\epsilon>0$. Kasami and Tokura \cite{KT70} characterized
the codewords in $RM(n,d)$ of weight up to twice the minimal
distance of the code (i.e up to distance $2^{1-d}$). Based on their
characterization one could conclude the following.
\begin{cor}[Corollary 10 in \cite{GKZ08}]\label{thm:gkz1}
$$
A(2^{1-d} - \epsilon) \le (1/\epsilon) ^ {2(n+1)}
$$
\end{cor}

Corollary~\ref{thm:gkz1} and simple lower bounds (which we show
later, see Lemma~\ref{lemma:lower_bound}) show that $A(\alpha) =
2^{\Theta(n)}$ for $\alpha \in [2^{-d},2^{1-d}-\epsilon]$ for any
$\epsilon>0$ (and constant $d$).

\subsection{Our Results}

Gopalan et al. \cite{GKZ08} left as an open problem the question of bounding the
list-decoding size of Reed-Muller codes beyond the minimal distance.
In particular, they ask what is the maximal $\alpha$ s.t. $L(\alpha) = 2^{O(n)}$.

In this work we answer their question. Specifically we show bounds
on the list-decoding size of Reed-Muller code for distances passing
the minimal distance. In fact, we show that the asymptotic behavior
of $L(\alpha)$, for all $0 \le \alpha \le 1$. Our first result shows that there exist "cut-off distances", at which
the list-decoding size changes from $2^{\Theta(n^{\ell})}$ to $2^{\Theta(n^{\ell+1})}$:

\begin{thm}[First main theorem - list-decoding size]\label{thm:main_list_decode_RM}
Let $1 \le \ell \le d-1$ be an integer, and let $\epsilon>0$.
For any $\alpha \in [2^{\ell-d-1},2^{\ell-d}-\epsilon]$
$$
L(\alpha) = 2^{\Theta(n^{\ell})}
$$
and $L(\alpha) = 2^{\Theta(n^d)}$ for any $\alpha \ge 1/2$.
\end{thm}

Using Theorem~\ref{thm:main_list_decode_RM}, and
Theorem~\ref{thm:gkz:global-alg} we obtain the following algorithmic
result for list-decoding Reed-Muller codes from an arbitrary
distance.

\begin{thm}[List-decoding algorithm]\label{thm:list-decode-alg}
Given a received word $R:\F_2^n \to \F_2$ that is at distance
$\alpha$ from $RM(n,d)$, for $\alpha \in
[2^{\ell-d-1},2^{\ell-d}-\epsilon]$. where $1 \le \ell \le d-1$ is
an integer, and $\epsilon>0$. There exists an algorithm that runs in
time $poly(2^{\Theta(n^{\ell})})$ and produces a list of all $p \in
RM(n,d)$ such that $dist(p,R) \leq \alpha$
\end{thm}

%
%
%

The weight distribution of $RM(n,d)$ codes beyond twice the minimum
distance was widely open prior to our work. See e.g. Research
Problem (15.1) in~\cite{MS} and the related discussion in that
Chapter.

In this work we provide asymptotic bounds for the weight
distribution of $RM(n,d)$ that applied for all weights $2^{-d} \le
\alpha \le 1/2$. Specifically, our second main result gives exact
boundaries on the range of $\alpha$ for which
$A(\alpha)=2^{\Theta(n^{\ell})}$, for any $\ell=1,2,...,d$.

\begin{thm}[Second main Theorem - accumulative weight distribution]\label{thm:main_weight_distrib_RM}
Let $1 \le \ell \le d-1$ be an integer, and let $\epsilon>0$.
For any $\alpha \in [2^{\ell-d-1},2^{\ell-d}-\epsilon]$
$$
A(\alpha) = 2^{\Theta(n^{\ell})}
$$
and $A(\alpha) = 2^{\Theta(n^d)}$ for any $\alpha \ge 1/2$.
\end{thm}

Theorems~\ref{thm:main_list_decode_RM} and~\ref{thm:main_weight_distrib_RM} are asymptotically tight
for constant $\epsilon>0$. For sub-constant $\epsilon$, and $\alpha \in [2^{\ell-d-1},2^{\ell-d}-\epsilon]$,
our bound gives:
$$
A(\alpha) \le L(\alpha) \le 2^{O(n^{\ell} / \epsilon^2)}
$$

We conjecture this dependency on $\epsilon$ is not optimal,
and the correct dependency should be $\log(1/\epsilon)$ instead of $1/\epsilon^2$. We expand more
on that in the body of the paper.

\subsection{Techniques}
The bounds on the accumulative weight distribution of the Reed-Muller code are obtained using the following novel strategy. We show
that a function $f:F_2^n \to F_2$ whose weight is bounded by $wt(f)
\leq 2^{-k}(1-\epsilon)$ can be {\em computed} as an expectation of
its $k$th-derivatives multiplied by some bounded coefficients
(Lemma~\ref{lemma:calc_by_ders}).

Using standard sampling methods we then show
(Lemma~\ref{lemma:calc_to_approx_by few}) that a function $f:F_2^n
\to F_2$ whose weight is bounded by $wt(f) \leq 2^{-k}(1-\epsilon)$
can be well approximated by a constant number $c=c(k,\epsilon)$ of its
$k$th-derivatives. This implies that every $RM(n,d)$ codeword of
weight up to $2^{-k}(1-\epsilon)$ can be well approximated by
$c=c(k,\epsilon)$ of its $k$th-derivatives. Since the distance
between every pair of $RM(n,d)$ codewords is at least $2^{-d}$, a
good enough approximation of a $RM(n,d)$ codeword determines the
Reed-Muller codeword uniquely. Hence, the
number of $RM(n,d)$ codewords up to weight $2^{-k}(1-\epsilon)$, is
bounded by the number of $k$th-derivatives to the power of
$c=c(k,\epsilon)$. As $RM(n,d)$ codewords are polynomials of degree
at most $d$, their $k$th-derivatives are polynomials of degree at
most $d-k$. There can be at most $\Theta(2^{n^{d-k}})$ such
derivatives. Thus, the number of $RM(n,d)$ codewords up to weight
$2^{-k}(1-\epsilon)$, can be bounded by $O(2^{n^{d-k}})^c = O(2^{c
\cdot n^{d-k}})$. We complement these upper bound estimations with
matching lower bounds.

A similar work in this line is the work of Viola and Bogdanov~\cite{BV},
which shows that a function
$f:F_2^n \to F_2$ whose weight is bounded by $wt(f) \leq
1/2-\epsilon$ can be well approximated by $c=c(k,\epsilon)$ of
its $1$st-derivatives. Note that approximation by $1$st-derivatives
{\em does not} imply in general approximation by $k$th-derivatives
which is crucial for obtaining our bounds here.

The bounds on the list-decoding size of Reed-Muller codes are
obtained using similar techniques to the ones used for bounding the
accumulative weight distributions.

\subsection{Generalized Reed-Muller Codes}
The problems of bounding both the accumulative weight distribution
and the list-decoding size can be extended to Generalized
Reed-Muller, the code of low-degree polynomials over larger fields.
However, our techniques fail to prove tight result in these cases.
We provide some partial results for this case and make a conjecture about the correct bounds
in Appendix~\ref{sec:GRM}.

\subsection{Organization}
Although our goal is bounding the list-decoding size of Reed-Muller codes,
we first study the accumulative weight distribution of Reed-Muller codes.
The techniques we develop are then easily transferred to bounding also the
list-decoding size.

The paper is organized as follows. In Section~\ref{sec:weight_RM} we
study the weight distribution of Reed-Muller codes and we prove the
Second Main Theorem (Theorem~\ref{thm:main_weight_distrib_RM}). In
Section~\ref{sec:list_RM} study the list-decoding size of Reed-Muller codes. We generalize
 the techniques of Section~\ref{sec:weight_RM} to prove the First Main Theorem
(Theorem~\ref{thm:main_list_decode_RM}). In Section~\ref{sec:GRM} we
study similar questions for Generalized Reed-Muller code and provide
non-tight bounds for these codes.






\section{Weight distribution of Reed-Muller codes}\label{sec:weight_RM}

In this section we study the weight distribution of Reed-Muller
codes, and we prove our Second Main Theorem
(Theorem~\ref{thm:main_weight_distrib_RM}). Let $RM(n,d)$ stand for
the code of multivariate polynomials $p(x_1,...,x_n)$ over $\F_2$ of
total degree at most $d$. In the following $n$ and $d$ will always
stand for the number of variables and the total degree. We will
assume that $d \ll n$, and study in particular the case of constant
$d$.

Our Second Main Theorem (Theorem~\ref{thm:main_weight_distrib_RM})
is a direct corollary of Theorem~\ref{thm:weight_distrib_RM}, giving
an upper bound on the accumulative weight at distance
$2^{\ell-d}-\epsilon$, and Lemma~\ref{lemma:lower_bound}, giving a
simple lower bound at distance $2^{\ell-d-1}$.

\begin{thm}[Upper bound on the accumulative weight]\label{thm:weight_distrib_RM}
For any integer $1 \le k \le d-1$,
$$
A(2^{-k}(1-\epsilon)) \le c_1 2^{c_2 \frac{n^{d-k}}{\epsilon^2}}
$$
where $c_1 = (1/\epsilon)^{O(d/\epsilon^2)}$ and $c_2 = O(d/(d-k)!)$. Importantly, $c_1,c_2$
are independent of $n$, and $c_2$ is independent of $\epsilon$. In particular for constant $d$ we get that
$$
A(2^{-k}-\epsilon) \le 2^{O(\frac{n^{d-k}}{\epsilon^2})}
$$
\end{thm}

\begin{lemma}[Lower bound on the accumulative weight]\label{lemma:lower_bound}
For any integer $1 \le k \le d$
$$
A(2^{-k}) \ge 2^{\frac{n^{d-k+1}}{(d-k+1)!}(1+o(1))}
$$
\end{lemma}

In the upper bound on $A(\alpha)$, while the dependence on $n$ is tight, we believe the dependence on $\epsilon$ can be improved.
For $k=d-1$ (and constant $d$), the characterization of \cite{KT70} shows that
$$
A(2^{1-d}-\epsilon) = 2^{\Theta(n \log(1/\epsilon))}
$$

We conjecture that this is the correct dependence on $\epsilon$ in all the range:
\begin{conj}
Let $d$ be constant. For any integer $1 \le k \le d-1$,
$$
A(2^{-k}-\epsilon) = 2^{\Theta(n^{d-k} \log(1/\epsilon))}
$$
\end{conj}

We start by proving the lower bound.
\begin{proof}[Proof of Lemma~\ref{lemma:lower_bound}]
Single out $k$ variables $x_1,...,x_k$, and let $q$ be any degree $d-k+1$ polynomials
on the remaining $n-k$ variables. First, for any such $q$, the following degree $d$ polynomial has
relative weight exactly $2^{-k}$:
$$
q'(x_1,...,x_n)=x_1 x_2 ... x_{k-1}(x_k + q(x_{k+1},...,x_n))
$$
The number of different polynomials $q$ is
$$
2^{{n-k \choose d-k+1}} = 2^{\frac{n^{d-k+1}}{(d-k+1)!}(1+o(1))}
$$
\end{proof}

We will prove Theorem~\ref{thm:weight_distrib_RM} in the rest of the section. We start by defining discrete derivatives,
which will be our main tool in the proof.

\begin{dfn}
Let $f:\F_2^n \to \F_2$ by a function. We define the discrete derivative of $f$ in direction $a \in \F_2^n$ to
be
$$
f_a(x) = f(x+a)+f(x)
$$

We define the iterated discrete derivative of $f$ in directions $a_1,...,a_k \in \F_2^n$ to be
$$
f_{a_1,...,a_k}(x) = (...((f_{a_1})_{a_2})...)_{a_k}(x) = \sum_{S \subseteq [k]} f(x + \sum_{i \in S} a_i)
$$
\end{dfn}

We note that usually derivatives are defined as $f_a(x)=f(x+a)-f(x)$, but since we are working over $\F_2$, we
can ignore the signs.

We define another notion which is central to our proof, namely the
bias of a function.

\begin{dfn}
The bias of a function $f:\F_2^n \to \F_2$ is
$$
bias(f) = \E_{x \in \F_2^n}[(-1)^{f(x)}] = \P[f=0]-\P[f=1] = 1 - 2wt(f)
$$
\end{dfn}

The following lemma will be the heart of our proof. It shows that if
a function $f$ has weight less than $2^{-k}$, then it can be
computed by a its iterated $k$-derivatives.

\begin{lemma}[Main technical lemma]\label{lemma:calc_by_ders}
Let $f:\F_2^n \to \F_2$ be a function s.t. $wt(f) < 2^{-k}(1-\epsilon)$. Then the function $(-1)^{f(x)}:\F_2^n \to \{-1,1\}$ can be written as
$$
(-1)^{f(x)} = \E_{a_1,...,a_k \in \F_2^n}[\alpha_{a_1,...,a_k} (-1)^{f_{a_1,...,a_k}(x)}]
$$
where $\alpha_{a_1,...,a_k}$ are real numbers, of absolute value of at most $\frac{10}{\epsilon}$
\end{lemma}

We will first prove Theorem~\ref{thm:weight_distrib_RM} given Lemma~\ref{lemma:calc_by_ders}, and then turn to prove Lemma~\ref{lemma:calc_by_ders}.
We will also need the following well-known technical lemma, which shows how to transform calculation by averaging many functions, to approximation
by averaging few functions.

\begin{lemma}[Approximation by sampling]\label{lemma:calc_to_approx_by few}
Let $f:\F_2^n \to \F_2$ be a function, $H=\{h_1,...,h_t\}$ a set of functions from $\F_2^n$ to $\F_2$, s.t. there exist
constants $c_{h_1},...,c_{h_t}$ of absolute value at most $C$, s.t.
$$
(-1)^{f(x)} = \E_{i \in [t]}[c_{h_i} (-1)^{h_i(x)}]\qquad(\forall x \in \F_2^n)
$$
Then $f$ can be approximated by a small number of the functions $h_1,...,h_t$. For any $\delta>0$, there exist
functions $h_1,...,h_{\ell} \in H$ for $\ell = O(C^2 \log{1/\delta})$, and a function $F:\F_2^{\ell} \to \F_2$, s.t. the
relative distance between $f(x)$ and $F(h_1(x),...,h_{\ell}(x))$ is at most $\delta$, i.e.
$$
\P_{x \in \F_2^n}[f(x) \ne F(h_1(x),...,h_{\ell}(x))] \le \delta
$$
The function $F$ is a weighted majority, i.e. it is of the form:
$$
F(h_1(x),...,h_{\ell}(x)) = sign(\frac{\sum_{i=1}^{\ell} s_i (-1)^{h_i(x)}}{\ell})
$$
where $sign(x)$ is defined by $sign(x)=1$ if $x \ge 0$ and $sign(x)=-1$ if $x<0$.
Moreover, we can have $s_1,...,s_{\ell}$ to be integers of absolute value at most $C+1$.
\end{lemma}

Using Lemmas~\ref{lemma:calc_by_ders} and~\ref{lemma:calc_to_approx_by few} we now prove Theorem~\ref{thm:weight_distrib_RM}.

\begin{proof}[Proof of Theorem~\ref{thm:weight_distrib_RM}]
Fix $1 \le k \le d-1$. We will bound the number of polynomials $p \in RM(n,d)$ s.t. $wt(p) \le 2^{-k}(1-\epsilon)$.
Let $p$ be any such polynomial. We apply Lemma~\ref{lemma:calc_by_ders} to $p$. We can write $(-1)^{p(x)}$ as
$$
(-1)^{p(x)} = \E_{a_1,...,a_k \in \F_2^n}[\alpha_{a_1,...,a_k} (-1)^{p_{a_1,...,a_k}(x)}]
$$
such that $|\alpha_{a_1,...,a_k}| \le \frac{10}{\epsilon}$.

We now apply Lemma~\ref{lemma:calc_to_approx_by few} to the set of polynomials $\{p_{a_1,...,a_k}(x): a_1,...,a_k \in \F_2^n\}$ with
$\delta = 2^{-(d+2)}$. We get that there are $\ell = O(\frac{d}{\epsilon^2})$ derivatives $\{p_{a^i_1,...,a^i_k}: i \in [\ell]\}$
s.t. the distance between $p(x)$ and $F(x)$ is at most $\delta$, where
$$
F(x) = sign(\frac{\sum_{i=1}^{\ell} s_i (-1)^{p_{a^i_1,...,a^i_k}(x)}}{\ell})
$$
and $s_1,...,s_{\ell}$ are integers of absolute value at most $O(\frac{1}{\epsilon})$.

We now make an important yet simple observation, that will let us bound the number of low weight polynomials
by bounding the number of functions $F(x)$. Given any $F(x)$, there can be at most one $p \in RM(n,d)$ s.t. $dist(F,p) \le \delta$.
Assume otherwise that there are two polynomials $p',p'' \in RM(n,d)$ s.t. $dist(p',F) \le \delta$ and $dist(p'',F) \le \delta$. By the triangle inequality $dist(p',p'') \le 2 \delta < 2^{-d}$, but this cannot hold if $p',p''$ are two different polynomials, since the
minimum relative distance of $RM(n,d)$ is $2^{-d}$.

So, if we bound the number of different functions $F(x)$ of the above form, we will also bound the number of polynomials $p$ of relative weight at most $2^{-k}(1-\epsilon)$. Consider the terms appearing in $F$:
\begin{itemize}
  \item We need $\ell = O(\frac{d}{\epsilon^2})$ derivatives and coefficients to describe $F$ completely.
  \item Any derivative $p_{a^i_1,...,a^i_k}(x)$ is a a polynomial of degree at most $d-k$, and so has at most $2^{{n \choose \le d-k}}$ possibilities.
  \item Any coefficient $s_i$ has $O(\frac{1}{\epsilon})$ possibilities.
\end{itemize}
Thus, the total the number of different $F$'s is at most
$$
\left( 2^{{n \choose \le d-k}} \cdot (1/\epsilon) \right)^{O(\frac{d}{\epsilon^2})} \le
c_1 2^{c_2 \frac{n^{d-k}}{\epsilon^2}}
$$
where $c_1 = (1/\epsilon)^{O(d/\epsilon^2)}$ and $c_2 = O(d/(d-k)!)$.

\end{proof}

We now turn to prove the Lemmas required for the proof of Theorem~\ref{thm:weight_distrib_RM}.
We prove Lemma~\ref{lemma:calc_by_ders} in Subsection~\ref{subsec:proof1} and
Lemma~\ref{lemma:calc_to_approx_by few} in Subsection~\ref{subsec:proof2}.

\subsection{Proof of the main technical lemma: Lemma~\ref{lemma:calc_by_ders}}\label{subsec:proof1}

Before proving Lemma~\ref{lemma:calc_by_ders}, we need some claims regarding derivatives.
The first claim shows that if a function has non-zero bias, it can be computed by an average
of its derivatives.

\begin{claim}\label{claim:calc_by_single_der}
Let $g:\F_2^n \to \F_2$ be a function s.t. $bias(g) \ne 0$. Then:
$$
(-1)^{g(x)} = \frac{1}{bias(g)} \E_{a \in \F_2^n}[(-1)^{g_a(x)}]
$$
where the identity holds for any $x \in \F_2^n$.
\end{claim}

\begin{proof}
Fix $x$. We have:
$$
(-1)^{g(x)} \E_{a \in \F_2^n}[(-1)^{g_a(x)}] = \E_{a \in \F_2^n}[(-1)^{g(x)-g_a(x)}] = \E_{a \in \F_2^n}[(-1)^{g(x+a)}] = bias(g)
$$
\end{proof}

The following claim shows that if a function has low weight, then derivatives of it will also have low weight,
and thus large bias.

\begin{claim}\label{claim:large_bias_for_der}
Let $f:\F_2^n \to \F_2$ be a function s.t. $wt(f) < 2^{-k}(1 - \epsilon)$.
Let $a_1,...,a_s \in \F_2^n$ for $1 \le s \le k-1$ be any derivatives, and consider $bias(f_{a_1,...,a_s})$. Then
$bias(f_{a_1,...,a_s}) \ge 1 - 2^{s+1-k}(1-\epsilon)$. In particular:
\begin{enumerate}
  \item If $s < k-1$ then $bias(f_{a_1,...,a_s}) \ge 1 - 2^{s+1-k}$
  \item If $s = k-1$ then $bias(f_{a_1,...,a_s}) \ge \epsilon$
\end{enumerate}
\end{claim}

\begin{proof}
Consider $f_{a_1,...,a_s}$
$$
f_{a_1,...,a_s} = \sum_{I \subseteq [s]} f(x + \sum_{i \in I} a_i)
$$

For random $x$, the probability that $f(x + \sum_{i \in I} a_i)=1$ is $wt(f)$, which is at most $2^{-k}(1-\epsilon)$.
Thus by union bound,
$$
\P_{x \in \F_2^n}[\exists I \subseteq [s],\ f(x + \sum_{i \in I} a_i)=1] \le 2^{s-k} (1 - \epsilon)
$$

In particular it implies that
$$
wt(f_{a_1,...,a_s}) = \P_{x \in \F_2^n}[f_{a_1,...,a_s}(x)=1] \le 2^{s-k} (1 - \epsilon)
$$

and we get the bound since $bias(f_{a_1,...,a_s}) = 1 - 2 wt(f_{a_1,...,a_s})$.
\end{proof}

We now can prove Lemma~\ref{lemma:calc_by_ders} using Claims~\ref{claim:calc_by_single_der} and~\ref{claim:large_bias_for_der}.

\begin{proof}[Proof of Lemma~\ref{lemma:calc_by_ders}]
Let $f:\F_2^n \to \F_2$ be a function s.t. $wt(f) \le 2^{-k}(1-\epsilon)$.
Thus $bias(f) = 1 - 2 wt(f) > 0$ and by Claim~\ref{claim:calc_by_single_der} we can write:
$$
(-1)^{f(x)} = \frac{1}{bias(f)} \E_{a_1 \in \F_2^n}[(-1)^{f_{a_1}(x)}]
$$

If $k=1$ we are done. Otherwise by Claim~\ref{claim:large_bias_for_der}, $f_{a_1}$ also has positive bias,
$$
bias(f_{a_1}) \ge 1 - 2^{s+1-k}(1-\epsilon) > 0
$$
and so again by Claim~\ref{claim:calc_by_single_der} we can write
$$
(-1)^{f_{a_1}(x)} = \frac{1}{bias(f_{a_1})} \E_{a_2 \in \F_2^n}[(-1)^{f_{a_1,a_2}(x)}]
$$

Thus we have:
$$
(-1)^{f(x)} = \frac{1}{bias(f)} \E_{a_1 \in \F_2^n}[\frac{1}{bias(f_{a_1})} \E_{a_2 \in \F_2^n}[(-1)^{f_{a_1,a_2}(x)}]]
$$

We can continue this process as long as we can guarantee that $f_{a_1,...,a_s}$ has non-zero bias for all $a_1,...,a_s \in \F_2^n$.
By Claim~\ref{claim:large_bias_for_der} we know this happens for $s \le k-1$, and thus we have:
$$
(-1)^{f(x)} = \E_{a_1,...,a_k \in \F_2^n}[\alpha_{a_1,...,a_k} (-1)^{f_{a_1,...,a_k}(x)}]
$$
where
$$
\alpha_{a_1,...,a_k} = \frac{1}{bias(f)} \frac{1}{bias(f_{a_1})} \frac{1}{bias(f_{a_1,a_2})} ... \frac{1}{bias(f_{a_1,...,a_{k-1}})}
$$

We now bound $\alpha_{a_1,...,a_k}$. By Claim~\ref{claim:large_bias_for_der} we get that:
$$
\alpha_{a_1,...,a_k} \le \frac{1}{\epsilon} \prod_{s=1}^{k-2} \frac{1}{1-2^{s-k+1}} \le \frac{1}{\epsilon} \prod_{r \ge 1} \frac{1}{1-2^{-r}} \le \frac{10}{\epsilon}
$$
\end{proof}

\subsection{Proof of Approximation by sampling Lemma: Lemma~\ref{lemma:calc_to_approx_by few}}\label{subsec:proof2}

\begin{proof}[Proof of Lemma~\ref{lemma:calc_to_approx_by few}]
Choose $h_1,...,h_{\ell}$ uniformly and independently from $H$. Fix $x \in \F_2^n$, and let $Z_i$ be the random variable
$$
Z_i = c_{h_i} (-1)^{h_i(x)}
$$

and let $S = \frac{Z_1 + ... + Z_{\ell}}{\ell}$.
We will use the fact that if $|S - (-1)^{f(x)}| < 1$ then $sign(S) = (-1)^{f(x)}$.

We first bound the probability that
$$
|S - (-1)^{f(x)}| > 1/4
$$

By regular Chernoff arguments for bounded independent variables,
since $\E[S] = (-1)^{f(x)}$ and each $Z_i$ is of absolute value of at most $C$, we get that
$$
\P_{h_1,...,h_{\ell} \in H}[|S - (-1)^{f(x)}| > 1/4] \le e^{-\frac{\ell}{32 C^2}}
$$
(see for example Theorem A.1.16 in \cite{AS00}).

In particular for $\ell = O(C^2 \log{1/\delta})$ we get that
$$
\P_{h_1,...,h_{\ell} \in H}[|S - (-1)^{f(x)}| > 1/4] \le \delta
$$

Thus by averaging arguments, there exists $h_1,...,h_{\ell}$ s.t.
$$
\P_{x \in \F_2^n}[|\frac{c_{h_1} (-1)^{h_1(x)} + ... + c_{h_{\ell}} (-1)^{h_{\ell}(x)}}{\ell} - (-1)^{f(x)}| \ge 1/4] \le \delta
$$

We now round each coefficient to a close rational, without damaging the approximation error. The coefficient
of $(-1)^{h_i(x)}$ is $\alpha_i = \frac{c_{h_i}}{\ell}$. If we round $c_{h_i}$ to the closest integer $[c_{h_i}]$, we get that
the coefficient of each $(-1)^{h_i(x)}$ is changed by at most $\frac{1}{2\ell}$, and thus the total approximation
is changed by at most $1/2$. Hence we have:
$$
\P_{x \in \F_2^n}[|\frac{[c_{h_1}] (-1)^{h_1(x)} + ... + [c_{h_{\ell}}] (-1)^{h_{\ell}(x)}}{\ell}) - (-1)^{f(x)}| \ge 3/4] \le \delta
$$

Thus we got that
$$
\P_{x \in \F_2^n}[sign(\frac{[c_{h_1}] (-1)^{h_1(x)} + ... + [c_{h_{\ell}}] (-1)^{h_{\ell}(x)}}{\ell}) \ne (-1)^{f(x)}] \le \delta
$$

\end{proof}

\section{List-decoding size of Reed-Muller codes}\label{sec:list_RM}
In this section we turn to the problem of bounding the list-decoding
size of Reed-Muller codes, and we prove the First Main Theorem
(Theorem~\ref{thm:main_list_decode_RM}). We will see that the same
techniques we used in Section~\ref{sec:weight_RM} to bound the
weight distribution, can be applied with minor variants to also
bound the list-decoding size.

%
%

The list-decoding size of a code is at least the accumulative weight
distribution, i.e. $L(\alpha) \ge A(\alpha)$. However, the
list-decoding size can sometimes be much larger than the
accumulative weight distribution.

%
%

Theorem~\ref{thm:main_list_decode_RM} is a direct corollary of
Theorem~\ref{thm:list_decode_RM}, giving an upper bound on the
list-decoding size at distance $2^{\ell-d}-\epsilon$, and the same
lower bound we used to bound the accumulative weight distribution,
obtained in Lemma~\ref{lemma:lower_bound}.

\begin{thm}[Upper bound on the list-decoding size]\label{thm:list_decode_RM}
For any integer $1 \le k \le d-1$,
$$
L(2^{-k}(1-\epsilon)) \le c_1 2^{c_2 \frac{n^{d-k}}{\epsilon^2} + c_3 \frac{n}{\epsilon^2}}
$$
where $c_1 = (1/\epsilon)^{O(d/\epsilon^2)}$, $c_2 = O(d/(d-k)!)$ and $c_3 = O(dk)$. Importantly, $c_1,c_2,c_3$
are independent of $n$, and $c_2,c_3$ are independent of $\epsilon$. In particular for constant $d$ we get that
$$
L(2^{-k}-\epsilon) \le 2^{O(\frac{n^{d-k}}{\epsilon^2})}
$$
\end{thm}

\begin{proof}[Proof of Theorem~\ref{thm:list_decode_RM}]
The proof will be similar to the proof of Theorem~\ref{thm:weight_distrib_RM}.
Fix $f:\F_2^n \to \F_2$ to be any function. We will bound the number of polynomials $p$ of degree at most $d$ s.t. $dist(p,f) \le 2^{-k}(1 - \epsilon)$.
Let $p \in RM(n,d)$ be such a polynomial, i.e. $dist(p,f) \le 2^{-k}(1 - \epsilon)$. Let $g(x) = p(x)-f(x)$, then
$wt(g) \le 2^{-k}(1 - \epsilon)$.
As in the proof of Theorem~\ref{thm:weight_distrib_RM}, we use the derivatives of $g$ to approximate $g$.
Set $\delta = 2^{-(d+2)}$. By Lemma~\ref{lemma:calc_by_ders} there are $\ell = O(\frac{d}{\epsilon^2})$ derivatives $\{g_{a^i_1,...,a^i_k}: i \in [\ell] \}$ s.t. the distance between $g(x)$ and $F(x)$ is at most $\delta$, where
$$
F(x) = sign(\frac{\sum_{i=1}^{\ell} s_i (-1)^{g_{a^i_1,...,a^i_k}(x)}}{\ell})
$$

Thus we have that $F+f$ approximates $p$, since:
$$
dist(p,F+f) = dist(p-f,F) \le \delta
$$

As in the proof of Theorem~\ref{thm:weight_distrib_RM}, given $F$ (and $f$) there can be at most a single $p \in RM(n,d)$
s.t. $dist(p,F+f) \le \delta$, and so if we will bound the number of functions $F$ we will bound the number of codewords close to $f$.

Consider the derivative $g_{a^i_1,...,a^i_k}(x)$ used in the expression for $F$. By linearity of derivation it can be decomposed as
$$
g_{a^i_1,...,a^i_k}(x) = p_{a^i_1,...,a^i_k}(x) - f_{a^i_1,...,a^i_k}(x)
$$

Each $p_{a^i_1,...,a^i_k}(x)$ is a degree $d-k$ polynomial, and so has at most $2^{{n \choose \le d-k}}$ possibilities.
Each $f_{a^i_1,...,a^i_k}(x) = \sum_{S \subseteq [k]}f(x + \sum_{j \in S}a^i_j)$ can be described by the values of
$a^i_1,...,a^i_k \in \F_2^n$, since we have access to $f$, and so has at most $2^{kn}$ possibilities. Each coefficient $s_i$ has $O(1/\epsilon)$
possibilities. Thus, in total the number of different $F$'s is at most
$$
\left( 2^{{n \choose \le d-k} + kn} \cdot (1/\epsilon) \right)^{O(\frac{d}{\epsilon^2})}
\le c_1 2^{c_2 \frac{n^{d-k}}{\epsilon^2} + c_3 \frac{n}{\epsilon^2}}
$$
where $c_1 = (1/\epsilon)^{O(d/\epsilon^2)}$, $c_2 = O(d/(d-k)!)$ and $c_3 = O(kd)$.
\end{proof}

{\em Acknowledgement.}
The second author would like to thank his advisor, Omer Reingold, for
on-going advice and encouragement. He would also like to thank Microsoft
Research for their support during his internship.

\appendix
\section{Generalized Reed-Muller codes}\label{sec:GRM}
The problems of bounding both the accumulative weight distribution
and the list-decoding size can be extended to Generalized
Reed-Muller, the code of low-degree polynomials over larger fields.
However, our techniques fail to prove tight result in these cases.
We briefly describe the reasons below, and give some partial
results.

We start by making some basic definitions. Let $q$ be a prime, and let $GRM_q(n,d)$ denote the code of multivariate polynomials $p(x_1,...,x_n)$ over the field $\F_q$, of total degree at most $d$.

\begin{dfn}
The relative weight of a function $f:\F_q^n \to \F_q$ is the fraction of non-zero elements,
$$
wt(f) = \frac{1}{q^n} |\{x \in \F_q^n: f(x) \ne 0\}|
$$
\end{dfn}

\begin{dfn}
The relative distance between two functions $f,g:\F_q^n \to \F_q$ is defined as
$$
dist(f,g) = \P_{x \in \F_q^n}[f(x) \ne g(x)]
$$
\end{dfn}

The accumulative weight distribution and the list-decoding size are defined analogously for $GRM_q(n,d)$, using
the appropriate definitions for relative weight and relative distance. We denote them by $A_q$ and $L_q$.
For each $1 \le k \le d$, we define a distance $r_k$:
\begin{enumerate}
  \item For $k=1$, let $d=(q-1)a + b$, where $1 \le b \le q-1$.
    Define $r_1 = q^{-a} (1-b/q)$.
  \item For $2 \le k \le d-1$, let $d-k=(q-1)a + b$, where $1 \le b \le q-1$.
    Define $r_k = q^{-a} (1-b/q) (1-1/q)$.
  \item For $k=d$, define $r_d = 1-1/q$.
\end{enumerate}

We conjecture that both for the accumulative weight distribution and the list-decoding size,
the distances $r_k$ are the thresholds for the exponential dependency in $n$:
\begin{conj}\label{conj:GRM_asymptotics}
Let $\epsilon>0$ be constant, and consider $GRM_q(n,d)$ for constant $d$. Then:
\begin{itemize}
  \item For $\alpha \le r_1 - \epsilon$ both $A_q(\alpha)$ and $L_q(\alpha)$ are constants.
  \item For $r_k \le \alpha \le r_{k+1}-\epsilon$ both $A_q(\alpha)$ and $L_q(\alpha)$ are $2^{\Theta(n^k)}$.
  \item For $\alpha \ge r_d$ both $A_q(\alpha)$ and $L_q(\alpha)$ are $2^{\Theta(n^d)}$.
\end{itemize}
\end{conj}

Proving lower bounds for $A_q(r_k)$ is similar to the case of $RM(n,d)$.
\begin{lemma}[Lower bound for $A_q$]\label{lemma:lower_bound_GRM}
For any integer $1 \le k \le d$,
$$
A_q(r_k) \ge 2^{\Omega(n^k)}
$$
\end{lemma}

The problem is proving matching upper bounds. Using directly the derivatives method we used to give upper bounds for $RM(n,d)$ gives the same bounds for $GRM_q(n,d)$, alas they are not tight for $q>2$:
$$
A_q(2^{-k}-\epsilon) \le 2^{O(n^{d-k})}
$$
If we would like to get upper bounds closer to the lower bounds, a natural approach would be to generalize  Lemma~\ref{lemma:calc_by_ders} to taking several derivatives in the same direction (which is possible
over larger fields). This would give us tight results for some values of $k$, if we could also
generalize Claim~\ref{claim:calc_by_single_der} to the case of taking a multiple derivative in the same direction.
However, we didn't find a way of doing so.

Instead, we give partial results for Conjecture~\ref{conj:GRM_asymptotics} in the two ends of the scale:
when $\alpha \le r_1 - \epsilon$, and when $r_{d-1} \le \alpha \le r_d - \epsilon$ (when $\alpha \ge r_d$ Lemma~\ref{lemma:lower_bound_GRM} gives $L_q(\alpha)$ and $A_q(\alpha)$ are both exponential in $n^d$).

First, the minimal distance of $GRM_q(n,d)$ is known to be $r_1$. Thus, for any $\epsilon > 0$, $A_q(r_1 - \epsilon) = 1$. Gopalan, Klivans and Zuckerman \cite{GKZ08} prove that $L_q(r_1 - \epsilon)$ is constant when
$q-1$ divides $d$:
\begin{thm}[Corollary 18 in \cite{GKZ08}]
Assume $q-1$ divides $d$. Then:
$$
L_q(r_1 - \epsilon) \le c(q,d,\epsilon)
$$
\end{thm}

Moving to the case of $r_{d-1} \le \alpha \le r_d - \epsilon$, we prove:
\begin{lemma}\label{lemma:upper_bound_GRM_r_d}
Let $\epsilon>0$ be constant. then:
$$
A_q(r_d - \epsilon) \le 2^{O(n^{d-1})}
$$
\end{lemma}

We now move on to prove Lemmas~\ref{lemma:lower_bound_GRM} and~\ref{lemma:upper_bound_GRM_r_d}.
We start with Lemma~\ref{lemma:lower_bound_GRM}:

\begin{proof}[Proof of Lemma~\ref{lemma:lower_bound_GRM}]
We start by proving for $2 \le k \le d-1$. Let $d-k=(q-1)a + b$, where $1 \le b \le q-1$.
Single out $a+2$ variables $x_1,...,x_{a+2}$, and let $g$ be any degree $k$ polynomial on the remaining variables. The following polynomial has degree $d$ and weight exactly $q^{-a} (1-b/q) (1-1/q)$:
$$
g'(x_1,...,x_n) = \left( \prod_{i=1}^{a} \prod_{j=1}^{q-1} (x_i-j) \right)
\left( \prod_{j=1}^{b} (x_{a+1}-j) \right)
\left(x_{a+2} + g(x_{a+3},...,x_n)\right)
$$
The number of distinct polynomial $g$ is $2^{\Omega(n^d)}$.

The proofs for $k=1$ and $k=d$ are similar: for $k=1$, let $d = (q-1)a+b$. Let $l_1(x),...,l_{a+1}(x)$ be any independent linear functions, and consider
$$
g'(x_1,...,x_n) = \left( \prod_{i=1}^{a} \prod_{j=1}^{q-1} (l_i(x)-j) \right)
\left( \prod_{j=1}^{b} (l_{a+1}(x)-j) \right)
$$
For $k=d$, let $g$ be any degree $d$ polynomial on variables $x_2,...,x_n$, and consider $g'(x_1,...,x_n) = x_1 + g(x_2,...,x_n)$.
\end{proof}

We now continue to prove Lemma~\ref{lemma:upper_bound_GRM_r_d}. We first make some necessary definitions.
\begin{dfn}
The bias of a polynomial $p(x_1,...,x_n)$ over $\F_q$ is defined to be
$$
bias(p) = \E_{x \in \F_q^n}[\omega^p(x)]
$$
where $\omega=e^{2 \pi i /q}$ is a primitive $q$-th root of unity.
\end{dfn}

Kaufman and Lovett \cite{KL08} prove that biased low-degree polynomials can be decomposed into a function of a constant number of lower degree polynomials:

\begin{thm}[Theorem 2 in~\cite{KL08}]\label{thm:bias_implies_low_rank}
Let $p(x_1,...,x_n)$ be a degree $d$ polynomial, s.t. $|bias(p)| \ge \epsilon$. Then $p$
can be decomposed as a function of a constant number of lower degree polynomials:
$$
p(x) = F(g_1(x),...,g_c(x))
$$
where $deg(g_i) \le d-1$, and $c=c(q,d,\epsilon)$.
\end{thm}

We will use Theorem~\ref{thm:bias_implies_low_rank} to bound $A(r_d - \epsilon)$ for any constant $\epsilon>0$.
\begin{proof}[Proof of Lemma~\ref{lemma:upper_bound_GRM_r_d}]
We will show that any polynomial $p \in GRM_q(n,d)$ s.t. $wt(p) \le 1-1/p - \epsilon$ can be decomposed as
$$
p(x) = F(g_1(x),...,g_c(x))
$$
where $deg(g_i) \le d-1$, and $c$ depends only on $q,d$ and $\epsilon$. Thus the number of such polynomials is bounded by the number of possibilities to choose $c$ degree $d-1$ polynomials, and a function $F:\F_q^c \to \F_q$. The number
of such possibilities is at most $2^{O(n^{d-1})}$.
Let $p$ be s.t. $wt(p) \le 1-1/p-\epsilon$. We will show there exists $\alpha \in \F_q$, $\alpha \ne 0$ s.t. $bias(\alpha p) \ge \epsilon$. We will then finish by using Theorem~\ref{thm:bias_implies_low_rank} on the polynomial
$\alpha p$.

Consider the bias of $\alpha p$ for random $\alpha \in \F_q$:
$$
\E_{\alpha \in \F_q}[bias(\alpha p)] = \E_{\alpha \in F_q, x \in \F_q^n}[\omega^{\alpha p(x)}] = 1 - wt(p)
$$
since for $x$'s for which $p(x)=0$, $\E_{\alpha \in F_q}[\omega^{\alpha p(x)}]=1$, and for $x$ s.t.
$p(x) \ne 0$, $\E_{\alpha \in F_q}[\omega^{\alpha p(x)}]=0$. We thus get that:
$$
\E_{\alpha \in \F_q \setminus \{0\} }[bias(\alpha p)] = 1 - \frac{q}{q-1} wt(p) \ge \frac{q}{q-1} \epsilon
$$
So, there must exist $\alpha \ne 0$ s.t. $bias(\alpha p) \ge \epsilon$.
\end{proof}

\end{document}